\documentclass[preprint,showpacs,preprintnumbers,amsmath,amssymb]{revtex4}

\usepackage{graphicx}
\usepackage{dcolumn}
\usepackage{bm}

\begin{document}


\title{Gain tuning and fidelity in continuous variable quantum teleportation}

\author{Toshiki Ide}
\affiliation{%
Department of Physics, Faculty of Science, University of Tokyo,\\
7-3-1 Hongo, Bunkyo-ku, Tokyo 113-0033, Japan 
}%
 \email{ide@femto.phys.s.u-tokyo.ac.jp}
\author{Holger F. Hofmann}%
\affiliation{%
CREST, Japan Science and Technology Corporation (JST),\\
Research Institute for Electronic Science, Hokkaido University, \\
Sapporo 060-0812, Japan
}%

\author{Akira Furusawa}
\affiliation{
Department of Applied Physics, \\ Faculty of Engeneering, University of Tokyo,\\
7-3-1 Hongo, Bunkyo-ku, Tokyo 113-8656, Japan}%

\author{Takayoshi Kobayashi}
\affiliation{%
Department of Physics, Faculty of Science, University of Tokyo,\\
7-3-1 Hongo, Bunkyo-ku, Tokyo 113-0033, Japan, \\
CREST, Japan Science and Technology Corporation (JST)}%

\begin{abstract}
The fidelity of continuous variable teleportation can be optimized by changing the gain in the modulation of the output field. 
We discuss the gain dependence of fidelity for coherent, vacuum and one photon inputs and propose optimal gain tuning strategies for corresponding input selections. 
\end{abstract}


\maketitle

\section{Introduction}

Continuous variable quantum teleportation transfers unknown quantum states of a light field input from Alice (sender) to Bob (receiver) using squeezed state entanglement as a resource \cite{Ben93,Vai94,Brau98,Fur98}. 
Since only finite squeezing is possible, the fidelity of this teleportation process is limited by non-maximal entanglement \cite{Brau98,Bra00,Hof00}. 
However, it has been shown in \cite{Hof00} that the 
output of a single pure state teleportation always
results in a pure state output conditioned by the
classical information sent from Alice to Bob. Therefore,
it may be possible that Bob can use this classical 
information to improve the fidelity if some information
on the selection of possible input states is known.

In the continuous variable quantum teleportation 
experiment realized by Furusawa et al. \cite{Fur98},  
the classical information is a complex field amplitude
$\beta$ that is effectively added to the output field
by a modulation process. The amplitude of this modulation
process can be modified by a gain factor. In \cite{Fur98},
the input state was a coherent state with an amplitude
much larger than one, such that the optimal fidelity was
obtained at a gain of one. However, Polkinghorne and Ralph
have pointed out that a lower gain can be useful for
teleporting photon entanglement \cite{Pol99}. 
Such considerations demonstrate that the optimal
gain for the teleportation depends on the selection of 
possible input states.

In the following, the dependence of fidelity on gain
is investigated for coherent states, for the vacuum
state, and for a one photon input. These results 
allow an optimization of fidelity for certain
groups of input states, such as coherent
states with constant amplitude and varying phase or
qubits of zero or one photon.

\section{Quantum state teleportation with variable gain}

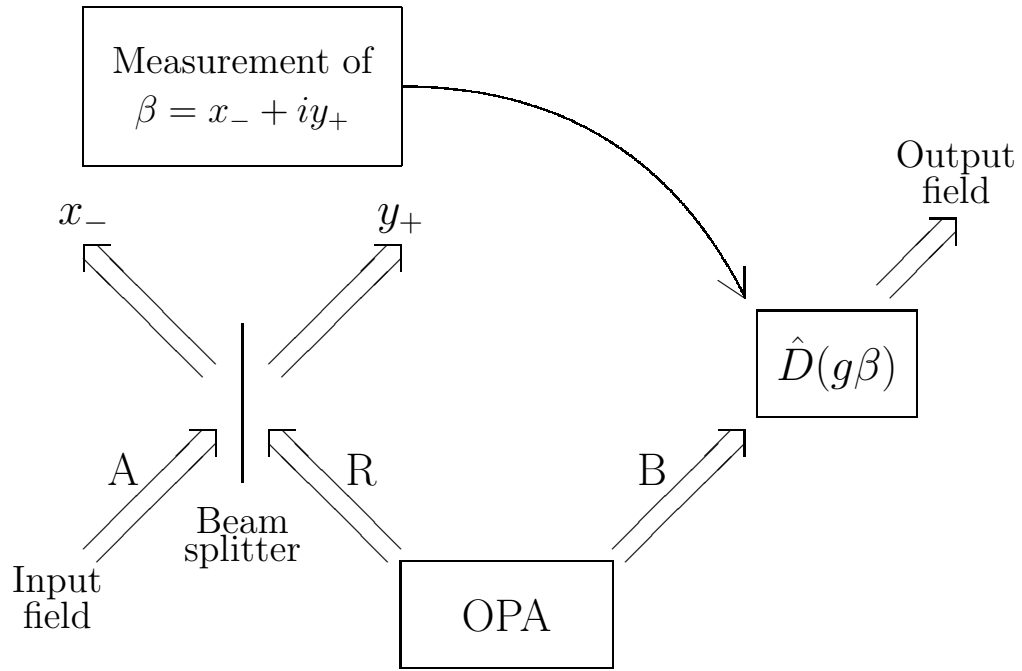
\begin{figure}
\begin{picture}(400,300)

\put(160,40){\framebox(80,40){\Large OPA}}

\put(160,85){\line(-1,1){45}}
\put(155,80){\line(-1,1){45}}
\put(110,130){\line(0,-1){10}}
\put(110,130){\line(1,0){10}}
\put(135,105){\makebox(20,20){\Large R}}

\put(240,85){\line(1,1){45}}
\put(245,80){\line(1,1){45}}
\put(290,130){\line(0,-1){10}}
\put(290,130){\line(-1,0){10}}
\put(245,105){\makebox(20,20){\Large B}}

\put(40,85){\line(1,1){45}}
\put(45,80){\line(1,1){45}}
\put(90,130){\line(0,-1){10}}
\put(90,130){\line(-1,0){10}}
\put(45,105){\makebox(20,20){\Large A}}
\put(10,66){\makebox(40,12){\large Input}}
\put(10,54){\makebox(40,12){\large field}}

\put(100,110){\line(0,1){60}}
\put(80,90){\makebox(40,12){\large Beam}} 
\put(80,78){\makebox(40,12){\large splitter}}

\put(90,155){\line(-1,1){45}}
\put(85,150){\line(-1,1){45}}
\put(40,200){\line(0,-1){10}}
\put(40,200){\line(1,0){10}}
\put(30,200){\makebox(20,20){\Large $x_-$}}

\put(110,155){\line(1,1){45}}
\put(115,150){\line(1,1){45}}
\put(160,200){\line(0,-1){10}}
\put(160,200){\line(-1,0){10}}
\put(150,200){\makebox(20,20){\Large $y_+$}}

\put(40,230){\framebox(120,60){}}
\put(60,260){\makebox(80,20){\large Measurement of}}
\put(60,240){\makebox(80,20){\large $\beta=x_-+i y_+$}}

\bezier{400}(160,260)(250,260)(290,180)
\put(290,180){\line(0,1){12}}
\put(290,180){\line(-3,2){10}}

\put(295,135){\framebox(60,40){\Large $\hat{D}(g\beta)$}}

\put(340,185){\line(1,1){25}}
\put(345,180){\line(1,1){25}}
\put(370,210){\line(0,-1){10}}
\put(370,210){\line(-1,0){10}}

\put(350,227){\makebox(40,12){\large Output}}
\put(350,215){\makebox(40,12){\large field}}
\end{picture}
\caption{\label{setup} Schematic representation of the quantum teleportation 
setup.}
\label{system}
\end{figure}

Fig.\ref{setup} shows the schematic sets of the quantum teleportation used in \cite{Fur98}.
Alice transmits an unknown quantum state $\mid \psi \rangle _{A}$ to Bob. Alice and Bob share EPR beams in advance. 
The quantum state of the EPR beams reads
\begin{equation}
\mid q \rangle_{R,B} = \sqrt{1-q^2} \sum_{n=0}^{\infty} q^n \mid n \rangle_{R} \mid n \rangle_{B},
\end{equation}
where R is the mode used by Alice as a quantum reference in the joint measurement of A and R \cite{Hof00}, and B is the output mode on Bob's side. The degree of entanglement is given by the parameter $q$. 
The parameter $q$ varies from 0 to 1, with $q=1$ for maximal entanglement and $q=0$ for no entanglement (vacuum in R and B).
Experimentally, $q$ is determined by the squeezing
achieved in the entangled modes. If the variance of
the squeezed quadratures is reduced by a factor of $\exp{(-2r)}$,
the entanglement is given by $q=\tanh (r)$.

Alice mixes her input state with reference EPR beam using a 50$\%$ beam-splitter and performs an entanglement measurement of the complex field value $\beta = x_{-}+iy_{+}$, where
\begin{eqnarray}
\hat{x}_{-}&=&\hat{x}_{A}-\hat{x}_{R}, \nonumber \\
\hat{y}_{+}&=&\hat{y}_{A}+\hat{y}_{R}.
\end{eqnarray}
As has been shown previously \cite{Hof00,Hof01}, the
output state of the teleportation process can then be
obtained by applying a transfer operator to the input,
\begin{equation}
\mid \psi_{\mbox{out}}(\beta) \rangle_{B} = \hat{T}^g_{q}(\beta) \mid \psi\rangle_{A}.
\label{trans}
\end{equation}
Note that the output state is not normalized, since $\langle \psi_{\mbox{out}}(\beta) \mid \psi_{\mbox{out}}(\beta) \rangle$ is the probability of obtaining the measurement result $\beta$ in the teleportation.

The transfer operator $\hat{T}^g_{q}(\beta)$ for variable
gain $g$ can be expressed using photon number states
$\mid n \rangle$ and displacement operators 
$\hat{D}(\beta)$,  
\begin{equation}
\hat{T}^{g}_{q}(\beta) = \sqrt{\frac{1-q^2}{\pi}}\sum_{n=0}^{\infty}q^n \hat{D}(g\beta) \mid n \rangle _{BA}\langle n \mid \hat{D}(-\beta).
\label{newtrans}
\end{equation}
This operator can be now be applied to various input 
states. 
In the case of a coherent state, the output state is 
\begin{eqnarray}
\hat{T}^{g}_{q}(\beta) \mid \alpha \rangle 
&=& 
\sqrt{\frac{1-q^2}{\pi}} e^{-(1-q^2) \frac{|\alpha-\beta|^2}{2}} 
e^{\frac{1}{2}(1-qg)(\alpha\beta^{\ast}-\alpha^{\ast}\beta)}
 \mid q\alpha+(g-q)\beta \rangle . \nonumber \\
 \label{gcohout}
\end{eqnarray}
This output state is also a coherent state with an amplitude given by a gain dependent superposition of $\alpha$ and $\beta$. Specifically, the gain factor affects the $\beta$ component of the output amplitude.
In the case of vacuum input state, the output state is 
\begin{eqnarray}
\hat{T}^{g}_{q}(\beta) \mid 0 \rangle 
&=& 
\sqrt{\frac{1-q^2}{\pi}} e^{-(1-q^2) \frac{|\beta|^2}{2}} 
 \mid (g-q)\beta \rangle . \nonumber \\ 
\label{g0out}
\end{eqnarray}
The vacuum is simply a coherent state with $\alpha=0$, so the output state is a coherent state with an amplitude proportional to $\beta$. The input vacuum state can be recovered by choosing $g=q$, effectively canceling the displacement \cite{Hof01}.
In the case of a one photon input state, the output state is 
\begin{eqnarray}
\hat{T}^{g}_{q}(\beta) \mid 1 \rangle 
&=& 
\sqrt{\frac{1-q^2}{\pi}} e^{-(1-q^2) \frac{|\beta|^2}{2}} 
\hat{D}((g-q)\beta)
\left ( (1-q^2) \beta^{\ast} \mid 0 \rangle + q \mid 1 \rangle \right ). \nonumber \\
\label{g1out}
\end{eqnarray}
This output state is a displaced quantum superposition of a vacuum component and a one photon component. Even though this is also a pure state conditioned by $\beta$, it is not possible to recover the one photon input state by varying the gain. 

\section{Fidelity and the effect of gain tuning}

The success or failure of quantum teleportation can be
characterized by the teleportation fidelity. It is defined by the
overlap between the input state and the output state
\cite{Bra00}. 
Using the results derived in the previous section, we can obtain
the gain factor dependence of teleportation fidelity for the
different input states.

\begin{figure}
\begin{picture}(350,400)
\put(40,0){\makebox(300,200){\includegraphics[width=10cm]{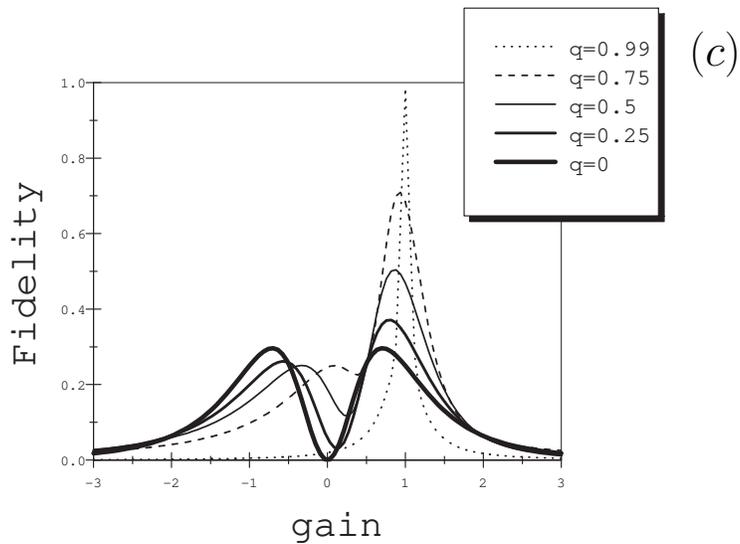}}}
\put(320,180){\makebox(40,10){\Large $(c)$}}
\end{picture}
\caption{The $g$ (gain factor) dependence of fidelity for a coherent states with $|\alpha|=1$ (a), the vacuum state (b) and the one photon state (c). The coherent state teleportation fidelity. The different curves correspond to entanglement parameter values of $q$=0.99 (dotted line),0.75 (dashed line),0.5 (thin line),0.25 (thick line),0 (thickest line), respectively. The peak positions shift to lower values of $g$ with decreasing $q$.}
\label{gfall}
\end{figure}

The total fidelity is obtained by averaging over all $\beta$,
even though $\beta$ is accessible classical information and the
output state is really a pure state conditioned by $\beta$. 
In this sense, gain tuning is a method to optimize the use of the
information $\beta$. 
In the case of a coherent state teleportation, the fidelity is 
\begin{eqnarray}
F_{q}^{\alpha}(g)
&=& 
\int d^2\beta
|\langle\alpha\mid\hat{T}^{g}_q(\beta)\mid\alpha\rangle|^2
\nonumber \\
&=&\int d^2\beta
e^{-(1-q^2)|\alpha-\beta|^2}e^{-|(1-q)\alpha-(g-q)\beta|^2}
\nonumber \\
&=& \frac{1-q^2}{1-2qg+g^2}\exp{\left(
-\frac{1-q^2}{1-2qg+g^2}(1-g)^2|\alpha|^2\right)}.
\nonumber \\
\label{fidcoh}
\end{eqnarray}
We note that maximal fidelity is always obtained at $g<1$. $g=1$
is optimal for $|\alpha|\rightarrow\infty$. 
Fig.(\ref{gfall}a) shows the gain dependence of the teleportation
fidelity for a coherent input state of amplitude 
$|\alpha|=1$ at different values of the entanglement parameter
$q$. 
The peak of the gain dependent fidelity shifts to lower
values of $g$ and gets lower and broader as the entanglement $q$
decreases. The peak is always at a gain value larger
than $g=q$, but lower than $g=1$. Therefore, gain tuning
to $g<1$ can improve the teleportation fidelity for coherent
input states. However, the fidelity remains below one
for all $q<1$. Moreover, the broadening of the peak indicates
that the fidelity is less sensitive to gain tuning for low values
of $q$.
In the special case of the vacuum state ($\alpha=0$), 
the fidelity is 
\begin{eqnarray}
F_{q}^{0}(g)
&=& 
\int d^2\beta |\langle 0 \mid\hat{T}^{g}_q(\beta)\mid 0
\rangle|^2 \nonumber \\
&=& \frac{1-q^2}{\pi}\int d^2\beta e^{-(1-2qg+g^2)|\beta|^2}
\nonumber \\
&=& \frac{1-q^2}{1-2qg+g^2} .
\label{fid0}
\end{eqnarray}
Fig.(\ref{gfall}b) shows the gain dependence of the teleportation
fidelity of the vacuum state at several values of the
entanglement parameter $q$. 
As in the general case of coherent states, the peak of the gain
dependent fidelity shifts to lower gain values and gets broader
as the entanglement parameter $q$ decreases. 
However, the peak value of the fidelity is always one at $g=q$.
The vacuum state is always teleported successfully at $g=q$
because this case corresponds to a simple attenuation at a beam
splitter with 
reflectivity $1-q^2$ \cite{Pol99,Hof01}.

In the case of a one photon state teleportation, the 
fidelity is
\begin{eqnarray}
F_{q}^{1}(g)
&=& 
\int d^2\beta |\langle 1 \mid\hat{T}^{g}_q(\beta)\mid 1
\rangle|^2 \nonumber \\
&=& \frac{1-q^2}{\pi}\int d^2\beta
e^{-(1-2qg+g^2)|\beta|^2}((1-qg)(g-q)|\beta|^2+q)^2 \nonumber \\
&=&
\frac{1-q^2}{(1-2qg+g^2)^3} 
( (g-q)^2 (1-qg)^2 + g^2(1-q^2)^2 )
.
\nonumber \\
\label{fid1}
\end{eqnarray}
Fig.(\ref{gfall}c) shows the gain dependence of the teleportation
fidelity of the one photon state at several values of the
entanglement parameter $q$. Two peaks appear when $q<1$. The
second peak is a result of the phase space symmetry of the single
photon input state. This can be understood most clearly at $q=0$,
where the output state is a coherent state $\mid g\beta \rangle$
and the overlap with the input photon number state is obviously
equal for $g$ and $-g$. 
The right peak of the fig.(\ref{gfall}c) changes in a similar way
to the peak of the fig.(\ref{gfall}a) for $|\alpha|=1$. 
Once again the maximal fidelity is always found at $g<1$. Both
peaks shift to lower gain values as the entanglement parameter
$q$ decreases. The right peak also gets broader and lower with
decreasing $q$.
The peak position is always between $g=q$ and $g=1$, 
with its lowest gain value at $g=1/\sqrt{2}$ for $q=0$.
Some additional gain is always necessary to replace the 
photon losses suffered at $g=q$ \cite{Hof01}.
Since the fidelity is always improved by lowering
the gain below one, we conclude that an optimal 
gain condition $g<1$ can be found for any selection
of states. In the following section, we apply this result 
to two examples.

\section{Optimal strategies for unknown input states}

\begin{figure}[h]
\begin{picture}(350,400)
\put(40,0){\makebox(300,200){\includegraphics[width=10cm]{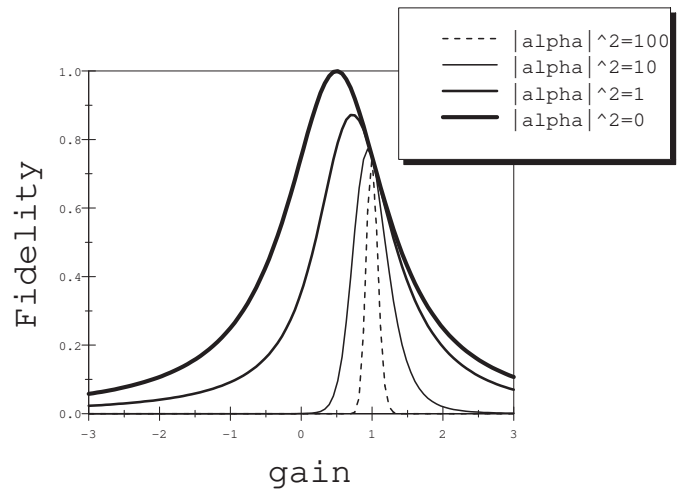}}}
\end{picture}
\caption{The gain dependence of the coherent state teleportation fidelity at an entanglement $q=0.5$. The curves correspond to input intensities $|\alpha|^2=0$ (thickest line),1 (thick line),10 (thin line),100 (dashed line). }
\label{fidamp}
\end{figure}

One possible selection of states to encode quantum information are coherent states with fixed amplitude and variable phase, 
$\mid \phi \rangle = \mid |\alpha| \exp[-i\phi]\rangle$.
For example, information could be encoded in the states
$\mid \alpha \rangle$,$\mid i \alpha \rangle$,
$\mid - \alpha \rangle$, and $\mid - i \alpha \rangle$.
Since the gain dependence of fidelity for coherent states does not depend on phase, the optimized gain is the same for all such states. It is therefore possible to determine the optimal gain directly from equation (\ref{fidcoh}).
Fig.\ref{fidamp} shows the gain $g$ dependence of the fidelity for input field intensities of $|\alpha|^2=0,1,10,100$ for teleportation with an entanglement of $q=0.5$. 
For $|\alpha|^2=0$ (vacuum input), the peak value of the fidelity is one at $g=q=0.5$. The input vacuum state is perfectly recovered by a gain tuning of $g=q=0.5$. 
With the increase of the input field intensity $|\alpha|^2$, the fidelity peak approaches 0.75 at $g=1$. For $|\alpha|^2>0$ the input coherent state cannot be recovered fully by gain tuning. 
Nevertheless some improvement of the fidelity is always possible through gain tuning to an optimal gain value of $q<g_{\mbox{opt}}(q)<1$.

\begin{figure}[h]
\begin{picture}(350,400)
\put(30,190){\makebox(70,20){\Large $g_{\mbox{opt}}$}}
\put(40,0){\makebox(300,200){\includegraphics[width=10cm]{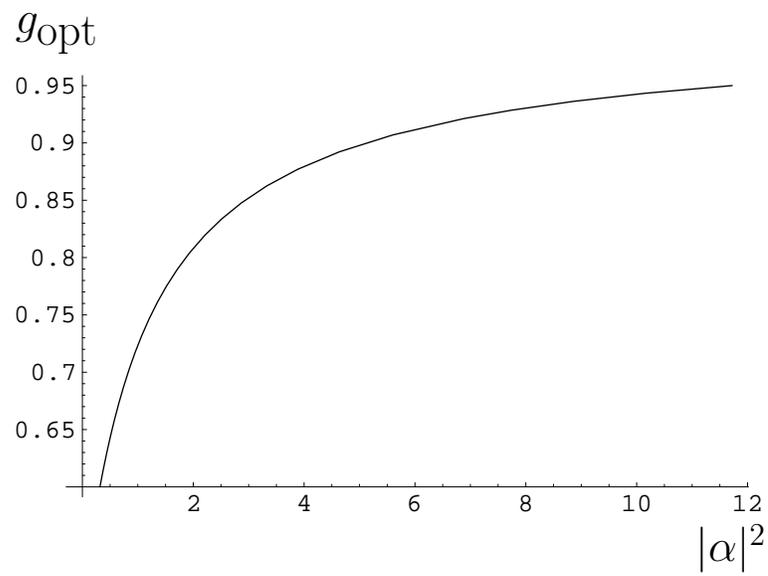}}}
\put(300,0){\makebox(40,10){\Large $|\alpha|^2$}}
\end{picture}
\caption{Relation between optimal gain and input intensity for a coherent state input at an entanglement parameter $q=0.5$.}
\label{optamp}
\end{figure}

\begin{figure}[h]
\begin{picture}(350,400)
\put(50,190){\makebox(70,20){\Large $F^{\alpha}(q)$}}
\put(40,0){\makebox(300,200){\includegraphics[width=10cm]{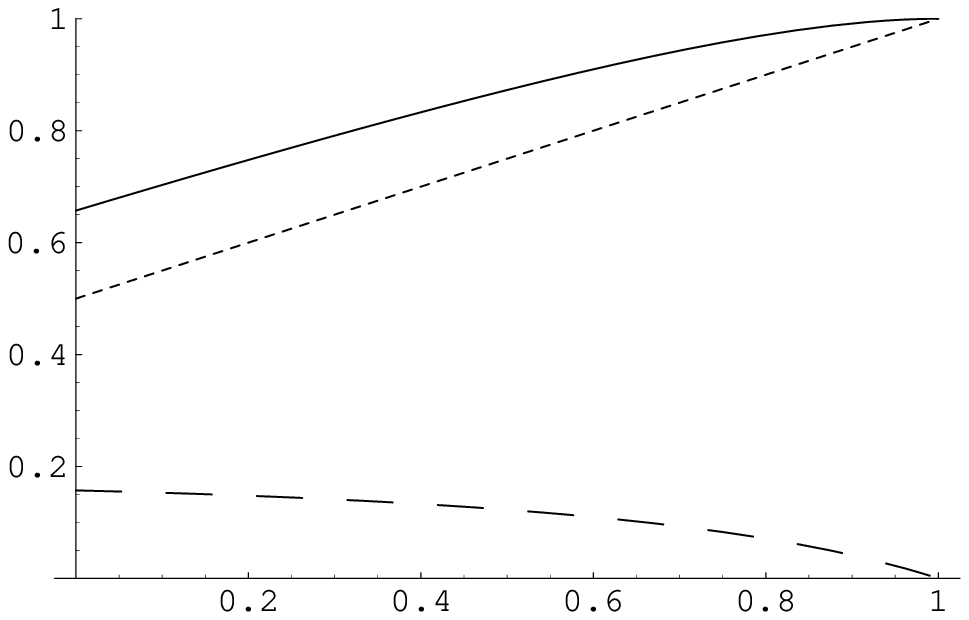}}}
\put(100,80){\makebox(40,10){\Large $F_{\mbox{non-opt}}$}}
\put(190,180){\makebox(40,10){\Large $F_{\mbox{opt}}$}}
\put(240,50){\makebox(40,10){\Large $\Delta F$}}
\put(300,0){\makebox(40,10){\Large $q$}}
\end{picture}
\caption{The solid line shows the optimized fidelity at $g_{\mbox{opt}}$ for the teleportation of a coherent state with $|\alpha|=1$. The dashed line shows the non-optimized fidelity at $g=1$ for comparison. The broken line shows the difference between $\Delta F=F_{\mbox{opt}}-F_{\mbox{non-opt}}$ between the two.}
\label{optfid}
\end{figure}

The optimal gain value is found by maximizing the fidelity. See the appendix for details of the calculation. For coherent states, the optimization condition depends on the intensity of the input fields. In its most compact form, it reads 
\begin{equation}
|\alpha|^2=\frac{(g_{\mbox{opt}}-q)
(1-2qg_{\mbox{opt}} +g_{\mbox{opt}}^2)}
{(1+q)(1-q)^2(1-g_{\mbox{opt}}^2)}.
\end{equation}
Figure \ref{optamp} shows this relation between $|\alpha|^2$
and $g_{\mbox{opt}}$ for an entanglement of $q=0.5$. 
The optimized gain $g_{\mbox{opt}}$ varies from $0.5(=q)$ 
at $|\alpha|^2=0$ to 1 for $|\alpha|^2 \to \infty$. 
The significance of gain tuning is already appreciable at
$|\alpha|^2 = 12$, where a gain of $g_{\mbox{opt}}=0.95$ is
optimal, and it rapidly approaches the vacuum situation
below intensities of $|\alpha|^2 = 4$. At $|\alpha|^2 = 1$,
the optimized gain is already as low as $g_{\mbox{opt}}=0.72$.
The improvement of fidelity achieved by gain tuning at
$|\alpha|^2 = 1$ is shown in fig.\ref{optfid}.
A substantial improvement of fidelity by gain tuning is observed
for almost all entanglement values. The increase in fidelity
achieved by gain tuning increases monotonously as the entanglement $q$
decreases, with a difference between the optimized and
non-optimized fidelity greater than 0.09 for $q < 0.7$. The
maximal increase in
fidelity is obtained in the limit of no entanglement ($q=0$)
with $\Delta F=0.16$. The fidelity for coherent state
teleportation
with known intensity can thus be significantly improved 
by an appropriate choice of the gain parameter
$g=g_{\mbox{opt}}$.
Note that the short analysis of gain tuning given in \cite{Fur98}
for the experimental realization of continuous variable
teleportation
was only applied to the high intensity limit of 
$|\alpha|^2 \to \infty$. As mentioned above, the maximal fidelity 
is then found to be extremely close to $g=1$. In this regime,
a very precise measurement is necessary to reveal the slight
shift 
of $g_{\mbox{opt}}$ due to the finite value of $|\alpha|$
actually used in the experiment. 

\begin{figure}[h]
\begin{picture}(350,400)
\put(150,190){\makebox(70,20){\Large $F_{\frac{1}{2}}(g)$}}
\put(40,0){\makebox(300,200){\includegraphics[width=10cm]{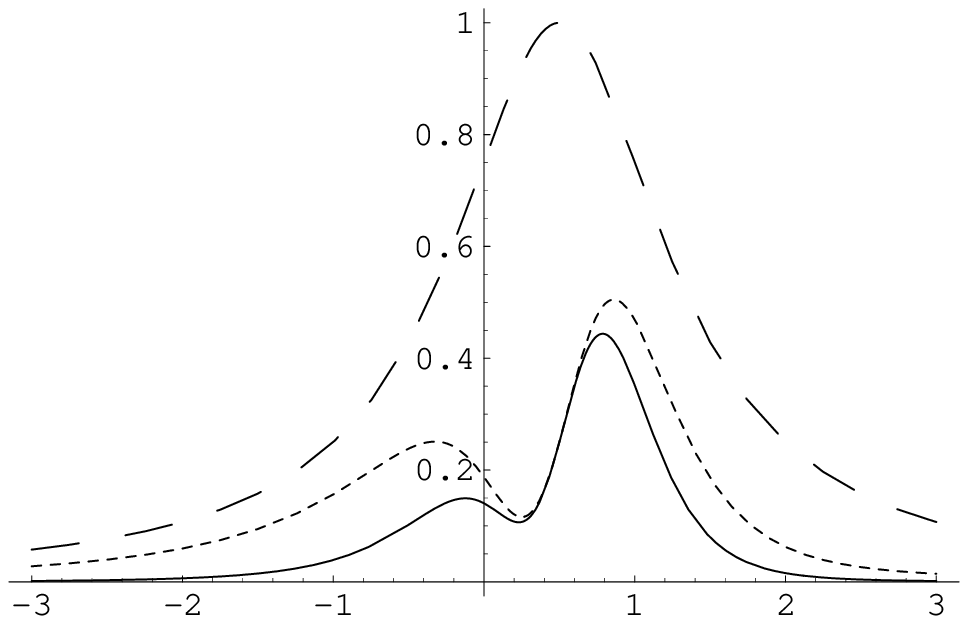}}}
\put(240,150){\makebox(40,10){\Large $F^0_{\frac{1}{2}}(g)$}}
\put(200,110){\makebox(40,10){\Large $F^1_{\frac{1}{2}}(g)$}}
\put(205,30){\makebox(40,10){\Large $F^{\mbox{joint}}$}}
\put(300,0){\makebox(40,10){\Large $g$}}
\end{picture}
\caption{The sloid line shows the joint fidelity of the two-mode
teleportation of a vacuum state and a single photon state as a
function of the gain factor $g$ at an entanglement of $q=0.5$.
The dashed lines show the fidelity of the single-mode
teleportation of vacuum state and of a single photon state,
respectively. The joint fidelity $F^{\mbox{joint}}$ is equal to
the product of $F^0_{\frac{1}{2}}(g)$ and
$F^1_{\frac{1}{2}}(g)$.}
\label{jointfid}
\end{figure}

Another typical encoding scheme for quantum information
uses the polarization states of single photons.
Continuous variable quantum teleportation can be applied
to such photonic qubits by teleporting each of two
orthogonal polarization modes in parallel \cite{Ide01}.
Since successful teleportation requires that both the
zero photon component and the one photon component of
the qubit are teleported without changes to the quantum
state, the total fidelity of the process can be
written as a product of the two individual fidelities
for vacuum and for single photon teleportation.
As shown in 
appendix B, this is even true if the polarization of
the qubit is unknown. Since neither the homodyne 
detection nor the displacement is sensitive to the
choice of polarization directions, the fidelity for
a single photon of unknown polarization is always
given by the joint fidelity $F^{\mbox{joint}}_q=F^0_q F^1_q$.
The gain dependence of this joint fidelity 
$F_q^{\mbox{joint}}=F_q^0F_q^1$ reads
\begin{equation}
F_{q}^{\mbox{joint}}(g)
= 
\frac{(1-q^2)^2}{(1-2qg+g^2)^4} 
( (g-q)^2 (1-qg)^2 + g^2(1-q^2)^2 )
.
\label{fidjoint}
\end{equation}
Fig.\ref{jointfid} shows the gain dependence of this fidelity
together with the fidelities for the vacuum and for the one
photon teleportation at an entanglement of $q=0.5$.
The main peak of the joint fidelity 
curve is found between the maxima of the vacuum and the 
single photon fidelities. 
At $q=0.5$, $g_{\mbox{opt}}=0.79$ gives a maximal joint fidelity
of 0.44, compared with a fidelity of 0.35 at $g=1$. 
The dependence of optimized gain on the entanglement parameter
$q$ can be determined by analytically maximizing the fidelity.
See the appendix for details of the calculation.  
Fig.\ref{jointgopt} shows the optimized gain 
$g_{\mbox{opt}}$ as a function of entanglement $q$ 
for both the photonic qubit and 
for a coherent state of intensity $|\alpha|^2=1$. 
Note that both curves are very close to each other,
suggesting that the gain tuning is quite similar for 
both single photons and coherent states with an average
photon number of one.
In the case of no entanglement at $q=0$, the 
optimized gain $g_{\mbox{opt}}(q)$ is 0.544 for the coherent 
state and 0.577 for the photonic qubit. It increases almost 
linearly to 1 as the entanglement $q$ is raised from 0 to 1. 
As a rule of thumb, optimal gain tuning for the teleportation
of photonic qubits is obtained at
\begin{equation}
g_{\mbox{opt}}(q) \approx 0.6 + 0.4 q.
\end{equation}
For practical purposes, this simplified relation should be
sufficient to achieve improved fidelities for single photon
teleportation. Note also that a similar optimization would
apply if the quantum information was encoded into vacuum or
one photon states within a single mode. 

\begin{figure}[h]
\begin{picture}(350,400)
\put(50,190){\makebox(70,20){\Large $F^{\mbox{joint}}(q)$}}
\put(40,0){\makebox(300,200){\includegraphics[width=10cm]{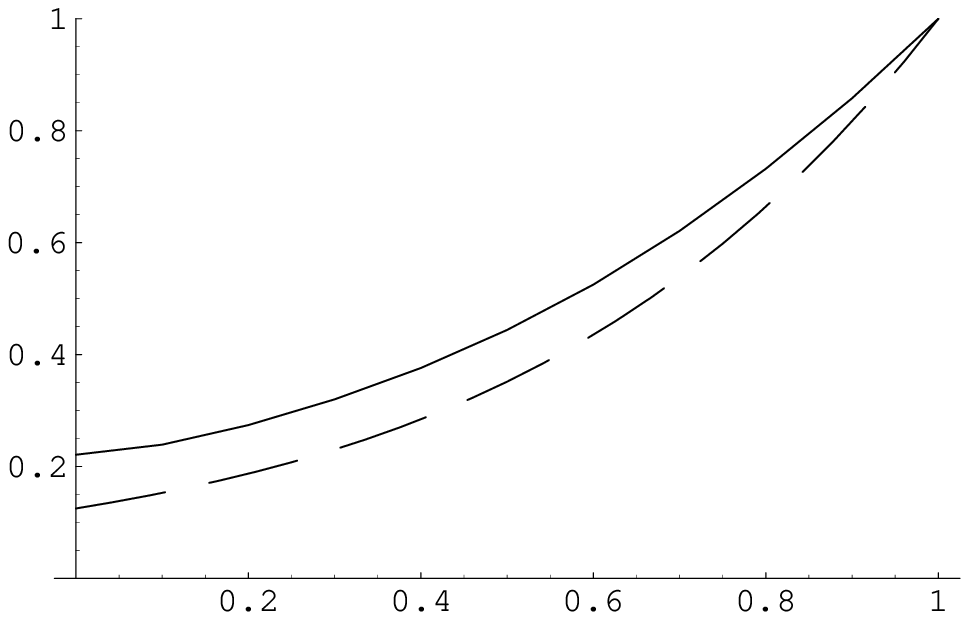}}}
\put(250,80){\makebox(40,10){\Large $F_{\mbox{non-opt}}$}}
\put(190,130){\makebox(40,10){\Large $F_{\mbox{opt}}$}}
\put(300,0){\makebox(40,10){\Large $q$}}
\end{picture}
\caption{The solid line shows the optimized fidelity at 
$g_{\mbox{opt}}$ for the teleportation of a photonic qubit. 
The dashed line shows the non-optimized fidelity at $g=1$ 
for comparison.}
\label{jointfidopt}
\end{figure}

The improvement of fidelity by optimized gain tuning for 
photonic qubit teleportation can be obtained from 
$g_{\mbox{opt}}(q)$ using equation (\ref{fidjoint}). 
Fig.\ref{jointfidopt} shows a comparison between
the optimized fidelity $F_{\mbox{opt}}$ and the 
non-optimized fidelity $F_{\mbox{non-opt}}$ as a function of
the entanglement parameter $q$. $F_{\mbox{opt}}$ is obtained with 
the optimized gain $g_{\mbox{opt}}(q)$, while for 
$F_{\mbox{non-opt}}$, the gain is fixed at $g=1$.
At $q=0$, the optimized fidelity is 0.221 while the non-optimized 
fidelity is 0.125. The difference of about 0.1 does not change
much up to $q=0.6$, so that an increase of 0.1 in fidelity is
possible for most cases of photonic qubit teleportation. Again,
this value is
similar to the improvement of fidelity achievable for coherent
states with $|\alpha|^2=1$. However, the improvement for photonic
qubits appears to be even more significant, given the relatively
low total fidelity of this teleportation.

\section{Conclusion}

The fidelity of continuous variable quantum teleportation can be
enhanced by varying the gain in the measurement dependent
modulation on the output field. We have shown that gain tuning
always maximizes the fidelity at a gain value of $g<1$. The
specific results for coherent states, vacuum, and single photons
have been obtained. Using these results,
the optimal gain tuning for the teleportation of coherent states
with known amplitude but unknown phase and for the teleportation
of the polarization of a single photon qubit
have been determined. For entanglement parameters of $q<0.7$, 
improvements of about $0.1$ are possible in the fidelity of 
single photon teleportation. Similar improvements are obtained
for
coherent states with intensity $|\alpha|^2=1$. These results
demonstrate the usefulness of gain tuning for input states with
a low average photon number.

\begin{acknowledgments}
We would like to thank Professor Mio Murao for helpful
discussions.
\end{acknowledgments}

\appendix

\section{Calculation for optimal gain}
Maximum fidelity for a coherent state teleportation is obtained
at a gain $g_{\mbox{opt}}$ which satisfies
$dF_q^{\alpha}(g)/dg=0$.
The derivative of the fidelity given by equation (\ref{fidcoh})
reads 
\begin{eqnarray}
\frac{dF_q^{\alpha}(g)}{dg}
&=&
2(-(g-q)(1-2qg+g^2)+(1+q)(1-q)^2(1-g^2)|\alpha|^2)
\nonumber \\
&&
\frac{1-q^2}{(1-2qg+g^2)^3}\exp{\left(
-\frac{1-q^2}{1-2qg+g^2}(1-q)^2|\alpha|^2  \right)}.
\label{diffid}
\end{eqnarray}
Therefore, the optimized gain $g_{\mbox{opt}}$ is given by the
polynomial
\begin{eqnarray}
g_{\mbox{opt}}^3-(a^2+3q)g_{\mbox{opt}}^2+
(1-2q^2)g_{\mbox{opt}}+(a^2-q)=0,
\label{poli}
\end{eqnarray}
where
\begin{eqnarray}
a^2&=&(1+q) (1-q)^2 |\alpha|^2.
\end{eqnarray}
The solution can be written as
\begin{eqnarray}
g_{\mbox{opt}}=\frac{1}{3}(A+\frac{B}{D}+D)
\label{gopt}
\end{eqnarray}
where
\begin{eqnarray}
A&=&(a^2+3q) , \nonumber \\
B&=&A^2+6q^2-3, \nonumber \\
C&=&A^3+27q^2-9a^2(2-q^2), \nonumber \\
D&=&(C+\sqrt{-B^3+C^2})^{1/3}.
\end{eqnarray}
We can obtain the optimized fidelity 
$F^{\alpha}_{\mbox{opt}}=F^{\alpha}(g_{\mbox{opt}})$ from
this result and equation(\ref{fidcoh}).

Likewise, the optimal gain for the teleportation of a photonic
qubit composed of a vacuum state and a single photon state
is obtained by differentiating equation (\ref{fidjoint}),
\begin{eqnarray}
\frac{dF_q^{\mbox{joint}}(g)}{dg}
&=&
\frac{dF_q^0(g)}{dg}F_q^1(g)+F_q^0(g)\frac{dF_q^1(g)}{dg}
\nonumber \\
&=&
-\frac{2(1-q^2)^2}{(1-2qg+g^2)^5}
(2 q^2 g^5 - 5 (q + q^3) g^4 + 2(3 + 3 q^2 + 4 q^4)g^3 \nonumber
\\
&&
- 4 q (2 + 2 q^2 + q^4) g^2 + 2(-1 + 4 q^2 + 2 q^4)g - 3 q^3 + q 
). 
\end{eqnarray}
Again, a polynomial for $g_{\mbox{opt}}$ is obtained. 
Since $g_{\mbox{opt}}$ must be one at $q=1$, we express this
polynomial in terms of the gain tuning parameter 
$h = 1-g_{\mbox{opt}}$. It then reads
\begin{eqnarray}
&&
2 q^2 h^5 + 5 q(1-q)^2 h^4 + 2(3 -10q + 13 q^2 -10 q^3 + 4
q^4)h^3 \nonumber \\
&&
-2 (1-q)^2 (9-q+8q^2-2q^3) h^2 
+ 4 (1-q)^2(4 -q + 3 q^2 -2 q^3) h \nonumber \\ 
&&
-4(1-q)^3 (1+q^2)
=0.
\label{jointgopt2}
\end{eqnarray}
where $h = 1-g_{\mbox{opt}}$. Solutions for this optimization
condition have been obtained numerically. 
As above, we can then determine the optimized fidelity 
$F^{\alpha}_{\mbox{opt}}=F^{\mbox{joint}}(g_{\mbox{opt}})$ from
this result and equation (\ref{fidjoint}).

\section{Polarization independence of the fidelity for
single photon qubit teleportation}

If a single photon of unknown polarization is teleported, 
the input photon state $\mid S \rangle $ is an unknown
superposition
of horizontal and vertical polarization states $\mid H \rangle=
\mid 1 
\rangle_H \mid 0 \rangle_V$ and $\mid V\rangle= \mid 0 \rangle_H
\mid 1 
\rangle_V$. The unknown polarization state can thus be written as
\begin{equation}
\mid S \rangle=c_H \mid H \rangle+c_V \mid V \rangle.
\end{equation}
Experimentally, this state is teleported by measuring 
$\beta_H$ and $\beta_V$ and applying the displacement
$D(\beta_H,\beta_V)$. In this process, the experimentalist
uses no information on the actual signal polarization S. 

It is nevertheless possible to transform the calculation 
of fidelity into the S,P basis, where P is the polarization 
orthogonal to S. 
If the unitary transform $\hat{U}$ rotates the polarization 
so that 
\begin{eqnarray}
\mid S \rangle = \hat{U} \mid H \rangle
\nonumber \\
\mid P \rangle = \hat{U} \mid V \rangle,
\end{eqnarray}
the fidelity for the teleportation of the $S$ polarized state can
be written as 
\begin{equation}
F^{joint}_q=\int d^2 \beta_H d^2 \beta_V 
|\langle H \mid \hat{U}^{\dagger} \hat{T}_q(\beta_H,\beta_V) 
\hat{U} \mid H \rangle |^2.
\end{equation}
It is now possible to apply the unitary transformations to
$\hat{T}_q(\beta_H,\beta_V)$. This is particularly simple 
for $\beta_H=\beta_V=0$, because $\hat{T}_q(0,0)$ is a function 
of the total photon number $n_{total}=n_H+n_V = n_S+n_P$, 
which is independent of the mode decomposition, 
\begin{eqnarray}
\hat{T}_{q}(0,0) 
&=&
\frac{1-q^2}{\pi} q^{(\hat{n}_H+\hat{n}_V)}
\nonumber \\
&=&
\frac{1-q^2}{\pi} q^{(\hat{n}_P+\hat{n}_S)}.
\end{eqnarray}
Therefore, $\hat{U}^\dagger\hat{T}_q(0,0)\hat{U}=\hat{T}_q(0,0)$.
The results for all other measurement values are obtained by
applying displacement operators to $\hat{T}_q(0,0)$. Since  
the displacement generated by $\hat{D}(\alpha_H,\alpha_V)$ is
linear 
in the field components, its transformation reads 
\begin{equation}
\hat{U}^{\dagger} \hat{D}(\alpha_H,\alpha_V) \hat{U} = 
\hat{D}(c_H \alpha_H + c_V \alpha_V,c_V \alpha_H - c_H 
\alpha_V).
\end{equation}
By combining these transformation properties, we obtain
\begin{equation}
\hat{U}^{\dagger} \hat{T}(\beta_H,\beta_V) \hat{U} = 
\hat{T}(c_H \beta_H + c_V \beta_V,c_V \beta_H - c_H 
\beta_V).
\end{equation}
It is therefore a straightforward matter to express the
transfer operator in the basis of the unknown input state,
even though this basis was not used in the experiment and
all measurement data was obtained in the H,V basis.
This property of the transfer operator greatly simplifies
the determination of the overall fidelity. By transforming the
integration using
\begin{eqnarray}
\beta_S=c_H \beta_H + c_V \beta_V
\nonumber \\
\beta_P=c_V \beta_H - c_H \beta_V,
\end{eqnarray}
the fidelity for the teleportation of an unknown polarization
reads
\begin{eqnarray}
F^{joint}_q
&=&
\int d^2 \beta_S d^2 \beta_P 
|\langle H \mid \hat{T}_q(\beta_S,\beta_P) \mid H \rangle |^2
\nonumber \\
&=&
\int d^2 \beta_S  
|\langle 1 \mid \hat{T}_q(\beta_S) \mid 1 \rangle |^2
\int d^2 \beta_P \; 
|\langle 0 \mid \hat{T}_q(\beta_P) \mid 0 \rangle |^2
.
\end{eqnarray}
Once more we would like to emphasize that this formulation of
fidelity does not depend on the polarization basis used in the
experiment and does not relate to any actual information
required for the teleportation process. 
It is therefore possible to apply the product of the single
photon fidelity and the vacuum fidelity to the teleportation 
of an unknown polarization state of a single photon.

\pagebreak

\begin{figure}
\begin{picture}(350,400)
\put(40,0){\makebox(300,200){\includegraphics[width=10cm]{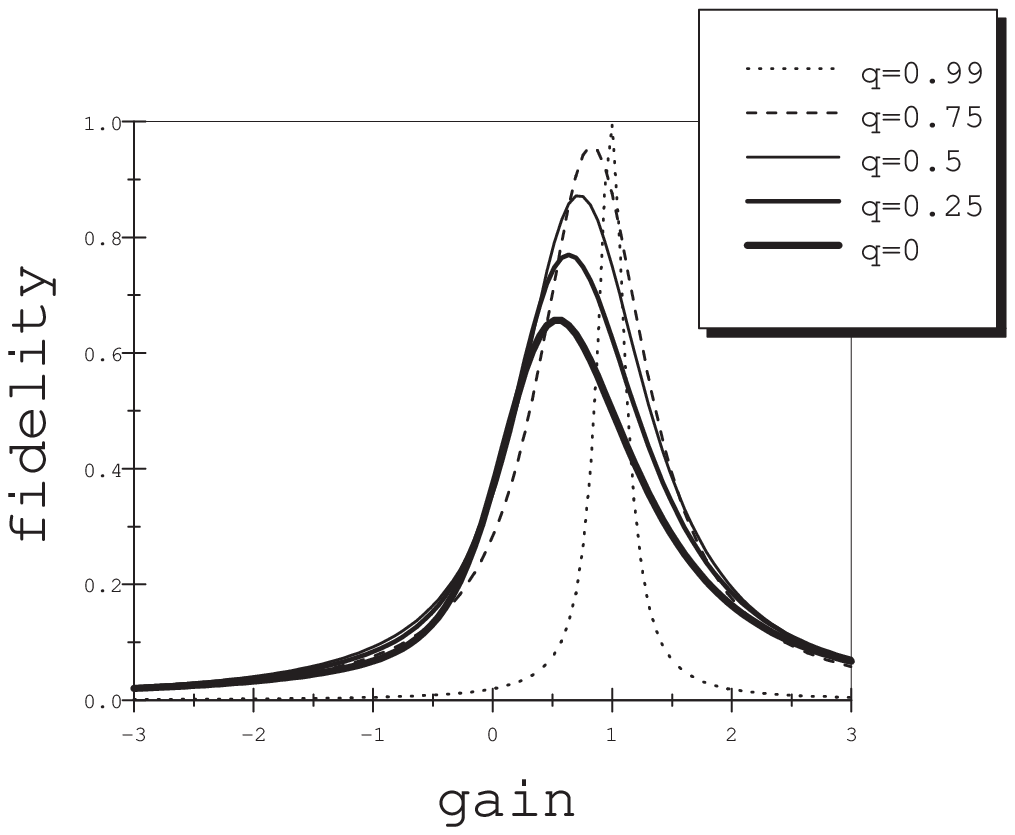}}}
\put(320,180){\makebox(40,10){\Large $(a)$}}
\end{picture}
\end{figure}

\begin{figure}[h]
\begin{picture}(350,400)
\put(40,0){\makebox(300,200){\includegraphics[width=10cm]{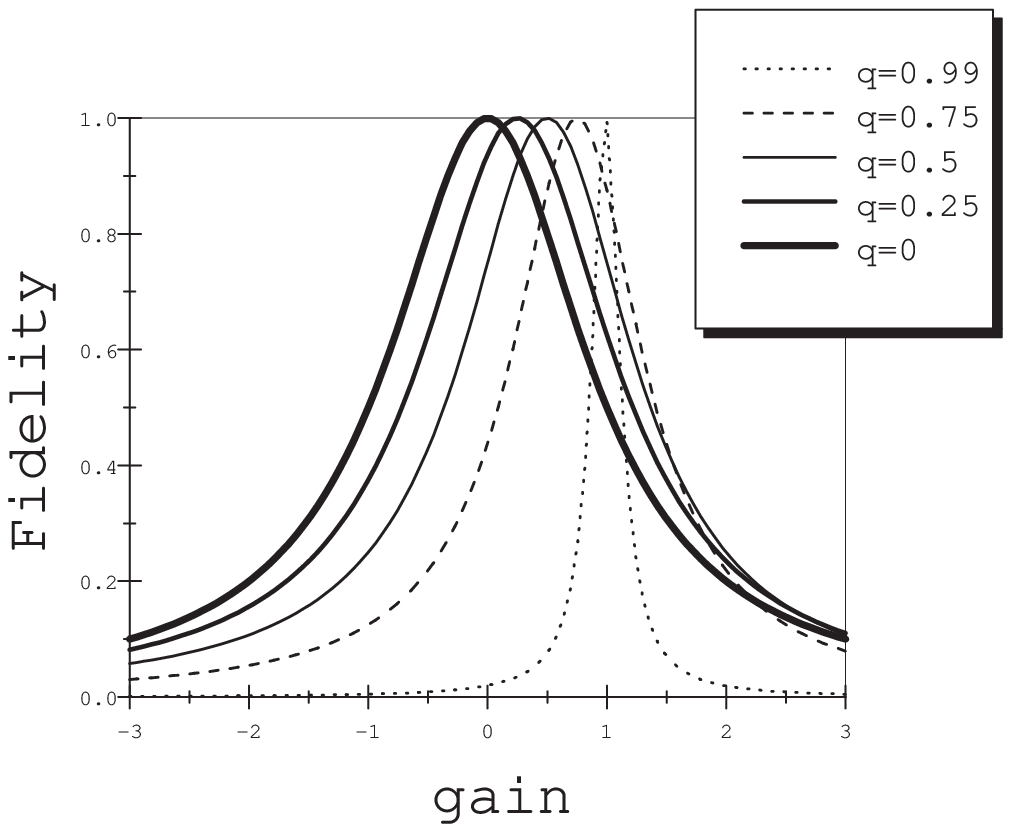}}}
\put(320,180){\makebox(40,10){\Large $(b)$}}
\end{picture}
\end{figure}

\begin{figure}[h]
\begin{picture}(350,400)
\put(50,190){\makebox(70,20){\Large $g_{\mbox{opt}}(q)$}}
\put(40,0){\makebox(300,200){\includegraphics[width=10cm]{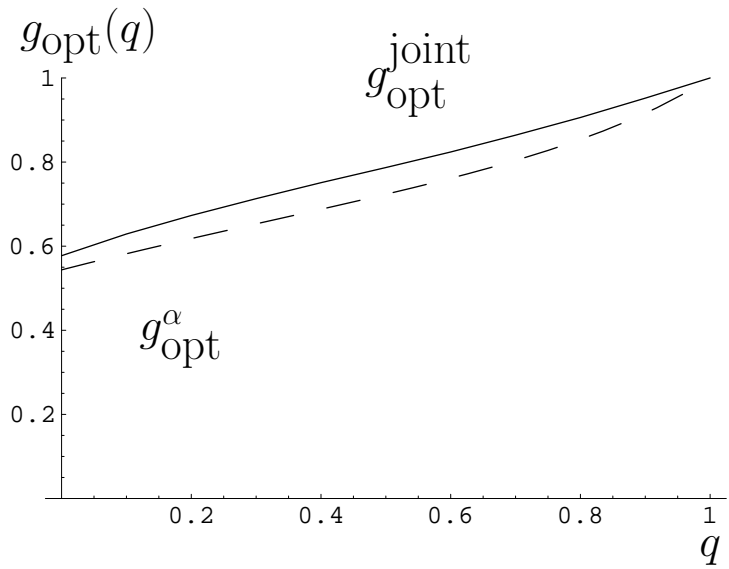}}}
\put(100,80){\makebox(40,10){\Large $g^{\alpha}_{\mbox{opt}}$}}
\put(190,180){\makebox(40,10){\Large $g^{\mbox{joint}}_{\mbox{opt}}$}}
\put(300,0){\makebox(40,10){\Large $q$}}
\end{picture}
\caption{Dependence of optimized gain $g_{\mbox{opt}}(q)$ on the
entanglement parameter $q$ for the teleportation of a photonic
qubit (solid line) and for the teleportation of a coherent state
with amplitude $|\alpha|=1$ (dashed line).}
\label{jointgopt}
\end{figure}

\end{document}